\documentclass[12pt,reqno]{article}
\usepackage[usenames]{color}
\usepackage{amssymb}
\usepackage{graphicx}
\usepackage{amscd}
\usepackage[colorlinks=true,
linkcolor=webgreen,
filecolor=webbrown,
citecolor=webgreen]{hyperref}
\definecolor{webgreen}{rgb}{0,.5,0}
\definecolor{webbrown}{rgb}{.6,0,0}
\usepackage{color}
\usepackage{fullpage}
\usepackage{graphics,amsmath,amssymb,amsthm}
\usepackage{amsfonts}
\usepackage{latexsym}
\usepackage{epsf}
\newtheorem{theorem}{Theorem}[section]

\newtheorem{lemma}[theorem]{Lemma}
\newtheorem{lem}[theorem]{Lemma}

\newcommand{\ff}{\mathfrak{f}}
\begin{document}
\begin{center}
{\Large On the problem of Frobenius in three numbers  }
\vskip 1cm
\large Abdelwaheb Miled\\
Universit\'e de Sousse - ISSAT\\
Cit\'e Ettaffala, Sousse 3000 - Tunisia\\
{\tt abdelwaheb.miled@issatso.rnu.tn}\\
\end{center}
\begin{abstract}
For three positive integers $a_{i},a_{j},a_{k}$ pairwise coprime, we present an algorithm that find the least multiple of  $a_{i}$ that is a positive linear combination of $a_j,a_{k}$. The average running time of this algorithm is $O(1)$. Using this algorithm and the chinese remainder theorem leads to a direct computation of the Frobenius number $\ff(a_1,a_2,a_3)$. 
\end{abstract}

\section{Introduction}

    We begin with some facts and notation \cite{Brauer,Johnson}. Let $a_1 < a_2 < a_3$ be relatively prime positive integers. It is well known that if $n$ is sufficientely large integer the Diophantine equation
   \begin{equation}
x_1a_1+x_2a_2+ x_3a_3=n
\end{equation}  
has solutions in positives integers $x_1,x_2,x_3$ and we say in this case that $n$ is $\mathbb{N}-$representable by $a_1,a_2,a_3$. Let $\ff(a_1,\cdots,a_n)$ be the largest integer not representable by the form $(1)$ in positive integers.\\
There are representation of $\ff(a_1,a_2,a_3)$ for each $i=1,2,3$ of the form  \cite{Johnson}: 
\begin{equation}
\ff(a_1,a_2,,a_3)={\sum}_{j\neq i}^{3}y_{ij}a_j \qquad y_{ij}>0
\end{equation}
If  $a_{1},a_{2},a_{3}$ are pairwise prime, we prove that in each representation above, there is only one term $y_{ij}a_j$ that is  $\mathbb{N}-$representable by $a_i,a_k$. \\
We give a simple algorithm that find the least multiple of $a_i$ that is representable by $a_j,a_k$. This method is very simple (few lines of code) and similar to euclidean algorithm. It compute $\ff(a_1,a_2,a_3)$ for large number, we test it for numbers  having thousand digits and the average running time is $O(1)$ for pairwise prime triplet of unmbers.\\ There are other algorithms to compute $\ff(a_1,a_2,a_3)$.  The Davison algorithm \cite{Davison} is based on the LLL lattice method. The elbows algorithm is \cite{Einstein}  is based on the integer linear programming  (ILP) method.\\
In section 2, we give semi-explicit representation of $\ff(a_{i},a_{j},a_{k})$ for  pairwise coprime integers $(a_{i},a_{j},a_{k})$. We present in Section 3 our algorithm for computing the least multiple $\mathbb{N}-$representable and prove that it terminates.\\
Section 4 is devoted to randomized testing and average time needed for computation of $\ff(a_1,a_2,a_3)$
\section{ Representation of $\ff(a_{i},a_{j},a_{k})$}
For any ordering $(i,j,k)$ of the set $\{1,2,3\}$, let $(a_{i},a_{j},a_{k})$ three integers relatively prime. It is no restriction to assume that $(a_{i},a_{j},a_{k})$ are pairwise coprime  and none of  $(a_{i},a_{j},a_{k})$ is $\mathbb{N}-$representable by the two others \cite{Brauer}.\\
\begin{lem}
\label{Lemma1}
Let 
\begin{equation}
\ff(a_{i},a_{j},a_{k})=x_j a_j+x_k  a_k, \qquad x_j,x_k>0
\end{equation}
 and suppose that
 \begin{equation}
  x_j  a_j={u}_{j} a_i+{v}_{j} a_k, \qquad {u}_{j} >0, {v}_{j} >0
  \end{equation}
  Then $x_j a_j$ is the smallest multiple of $a_j$ that is $\mathbb{N}-$representable by $a_i,a_k$
\begin{proof}
Suppose that $\ff(a_{i},a_{j},a_{k})$ have another representation: 
\begin{equation}
\ff(a_1,a_2,a_3)=y_j a_j+y_k a_k, \qquad y_j>x_j,z_j>0
\end{equation}
And
\begin{equation}
y_j a_j={s}_{j} a_i+{t}_{j} a_k, \qquad s_{j},t_{j}>0
\end{equation}
Combining the two expressions, we can represent $\ff(a_{i},a_{j},a_{k})$ as:
\begin{equation}
 \ff(a_{i},a_{j},a_{k})= u_{j} a_{j}+(y_j-x_j)a_j+(v_{j} +{t}_{k})a_k,  {u}_{j} >0, \qquad ({v}_{j} +{t}_{k})>0,{v}_{j} +{t}_{k}>0
 \end{equation}
so $\ff(a_{i},a_{j},a_{k})$ would be representable by $a_1,a_2,a_3$, a contradiction to the definition of $\ff(a_{i},a_{j},a_{k})$. 
\end{proof}
\end{lem}

\begin{lem}
\label{lem2}
For each $i$ let $\ff(a_{i},a_{j},a_{k})=y_{ij}a_j+y_{ik}a_k$ with $(i,j,k)$ a permutation cycle of $(1,2,3)$. Then either $y_{ij}a_j$ is representable by $(a_{i},a_{k})$, or $y_{ik}a_k$ representable by $(a_{i},a_{j})$
 \begin{proof}
Reresent $\ff(a_{i},a_{j},a_{k})$ as:
\begin{equation}
\ff(a_1,a_2,a_3)=y_{12}a_2+y_{13}a_3=y_{21}a_1+y_{23}a_3=y_{31}a_1+y_{32}a_2, \qquad y_{ij}>0, i,j=1,2,3
\end{equation}
Consider the couple $(y_{13},y_{23})$, observe first that $y_{13} \neq y_{23}$, otherwise $y_{12}a_2=y_{21}a_1$ a contradiction to $a_{1},a_{2}$ relatively prime.\\
If $y_{13} < y_{23}$ then
\begin{equation}
y_{12}a_2=y_{21}a_1+(y_{23}-y_{13})a_3, \qquad y_{12}>0,(y_{23}-y_{13})>0
\end{equation}
So $y_{12}a_2$ is representable by $(a_{1},a_{3})$.
Alternatively the case $y_{13} > y_{23}$ lead to
\begin{equation}
y_{21}a_1=y_{12}a_2+(y_{13}-y_{23})a_3,  \qquad y_{13}>0,(y_{13}-y_{23})>0
\end{equation}
And then $y_{21}a_1$ is representable by $a_2,a_3$.\\
We obtain the other results by considering the couples $(y_{12},y_{32})$ and $(y_{13},y_{31})$
\end{proof}
\end{lem}

Consider now integers $n$ that is not $\mathbb{N}-$representable by $a_1,a_2,a_3$ but is $\mathbb{N}-$representable by any couples $(a_j,a_k)$:
\begin{equation}
n=u_{12}a_{2}+u_{12}a_{2}=u_{21}a_{1}+u_{23}a_{3} =u_{31}a_{1}+u_{32}a_{2}\qquad u_{ij} \in \mathbb{N}^{*}
\end{equation}
In the first right expression above, suppose $u_{12}a_2$  $\mathbb{N}-$representable by $a_{1},a_{3}$. Lemma \ref{lem2},  implies that the $\mathbb{N}-$representable  term for the third expressions must be $u_{31}a_1$, and consequently the term $\mathbb{N}-$representable for the second expression $u_{21}a_1$. This give us a first decomposition of $n$. An other decomposition is given when we suppose  $u_{13}a_3$  $\mathbb{N}-$representable in the first right expression. Finally, such integers $n$ have exactly two possible decompositions.\\

To compute  $\ff(a,b,c)$ consider the two congruent systems:\\
\begin{minipage}[t]{.4\linewidth}
$\left\{\begin{array}{c} x \equiv u_{12}a_{1} \qquad \operatorname{mod} a_{2} \\ x \equiv u_{23}a_{2} \qquad \operatorname{mod} a_{3} \\x  \equiv u_{31}a_{3} \qquad \operatorname{mod} a_{1}\end{array}\right.$
 \end{minipage}
$ \left\{\begin{array}{c}x \equiv u_{21}a_{2} \qquad \operatorname{mod} a_{1} \\x \equiv u_{13}a_{1} \qquad \operatorname{mod} a_{3} \\x \equiv u_{32}a_{3} \qquad \operatorname{mod} a_{2}\end{array}\right.$
  \\
Solutions of the two congruent systems are given by the chinese remainder theorem and $\ff(a,b,c)$ is  maximum of the two solutions.  
\section{Least multiple of $b$ $\mathbb{N}-$representable by $a,c$}
\textbf{Algorithm} (Find least multiple representable).We are given three integers  $a,b,c$ pairwise prime. This algorithm find  the least multiple of $b$ that is $\mathbb{N}-$representable by $(a,c)$. 
\begin{enumerate}
\item{} [Initialize]\\
\qquad $p_{0}=\dfrac{1}{a}\big(b+c\bigl(\dfrac{-b}{c} \mod a \bigr) \bigr)$\\
\\
 $b \equiv p_{0}a \mod c$
\item{}[Sequences chains] 

Develop the integer sequences $(k_{i}),(p_{i},(v_{i})$, where\\
\begin{equation}
k_{1}=1+\left\lfloor \dfrac{c}{p_{0}}  \right\rfloor, \qquad k_{i}=1+\left\lfloor \dfrac{p_{i-2}}{p_{i-1}}\right\rfloor,  i \geqslant 2
\end{equation}
\begin{equation}
\label{suitev1}
p_{1}=k_{1}p_{0} \mod c, \qquad p_{i}\equiv k_{i}p_{i-1}  \operatorname{mod}p_{i-2}, i \geqslant 2 
\end{equation}
\begin{equation}
\label{suitev}
v_{i}=p_{i}\bigl(\dfrac{1}{p_{0}} \operatorname{mod} c\bigr)\operatorname{mod} c, \qquad i\geqslant 1
\end{equation}
\item{}[Main loop]\\
\\
While($p_{i}a>v_{i}b$) $\{$\\
\qquad\qquad $i=i+1$\\
\qquad $\}$\\
return $v_{i}a$
  \end{enumerate}
  The theoretical justification of this algorithm is as follows:
\begin{lemma}
The integer sequence $(p_{i})$ defined by equation \ref{suitev}  is strictly decreasing
\begin{proof}
For $i=1$, the integer $k_{1}$ defined by \ref{suitev} is the smallest $k$ verifying: $kp_{0}>c$.
In fact, we have from \ref{suitev1}: $p_{1}=k_{1}p_{0} \mod c$, so necessarily  $p_{1}<p_{0}$, otherwise $(k_{1}-1)$ would be the least integer $k$ satisfying $kp_{0}>c$.\\
For  $i \geqslant 2$, the integer $k_{i}$ defined by \ref{suitev} is the least integer $k$ such that $kp_{i-2}>p_{i-1}$ and with  $p_{i}\equiv k_{i}p_{i-1}  \operatorname{mod}p_{i-2}$,necessarily  $p_{i}<p_{i-1}$, otherwise $(k_{i}-1)$ would be the least integer $k$ satisfying $kp_{i-2}>p_{i-1}$.\\
\end{proof}
\end{lemma}
 
\section{Example}
Let $a = 7523,b = 8231,c = 9533$\\
\begin{table}[htdp]
\begin{center}\begin{tabular}{|c|c|c|c|c|}
\hline Step & $k_{i}$&$v_{i}$ & $p_{i}$ & $\dfrac{p_{i}a-v_{i}b}{c}$ \\\hline
\hline 1 & 2&9533 & 7001 & -3525 \\\hline2&2 & 7001 & 4469 & -1526 \\\hline 3& 3&4469 & 1937 & -1053 \\\hline 4 &2 & 1937 & 1342 & -580 \\\hline 5 & 2& 1342 & 747 & -107 \\\hline 6 & 5& 747 & 152 & 45 \\\hline \end{tabular} \caption{The least multiple of $b$ representable by $a,c$ is $747b=152c+45a$}
\end{center}
\label{defaulttable}
\end{table}\\
We test the number of step required to compute the three least multiple representable.\\
We use always the scheme $xb=\alpha c - \beta a, ya =\eta c - \nu b, zc=\mu b- \gamma a $\\ 
An implemented demonstration of this algorithm using Mathematica software is available  in \cite{Miled}.\\ 
Compared to Davison algorithm  \cite{Davison} we noted that our algorithm don't need to go through the entire euclidean chain until reaching zero but rather the first $p_{i}$ smaller than $v_{i}$. This seems giving substantial improvement of the running time.   

\bigskip
\hrule
\bigskip
\noindent 2000 {\it Mathematics Subject Classification}:
Primary 05A15; Secondary 52C07.\\
\noindent \emph{Keywords: } \\
Frobenius problem, Chinese remainder theorem, $\mathbb{N}-$representable numbers.

\begin{thebibliography}{Niren}

\bibitem[1]{Beihoffer}Dale Beihoffer, Jemimah Hendry, Albert Nijenhuis, and Stan Wagon,
Faster algorithms for Frobenius numbers, {\it Electron. J. Combin.} {\bf12} (2005).

\bibitem[2]{Brauer}A. Brauer,J. Shockley On a problem of Frobenius, {\it J. Reine Angew. Math.}  {\bf 211} (1962), 215--220.

\bibitem[3]{Davison}J. Davison. On the linear Diophantine Problem of Frobenius  {\it. J. of Number Theory} {\bf 48} (1994) 353Ð363.

\bibitem[4]{Einstein}D. Einstein, D. Lichtblau, A. Strzebonski, S. Wagon, Frobenius numbers by lattice point enumeration,{\it INTEGERS: Electron. J.  of Combinatorial Number Theory} {\bf7} (2007)

\bibitem[5]{Johnson}S. M. Johnson,  A linear diophantine problem, {\it Canad. J. Math.} {\bf 12} (1960), 390--398.

\bibitem[6]{Miled}A. Miled,\\  {\tt http://demonstrations.wolfram.com/PositiveFrobeniusNumbersOfThreeArguments}

\end{thebibliography}
\end{document}